\documentclass[twocolumn,aps,superscriptaddress,longbibliography]{revtex4-1}
\usepackage{palatino}
\usepackage[latin1]{inputenc}
\usepackage{epsf}
\usepackage{amsmath,amssymb}
\usepackage{latexsym}
\usepackage{calc}

\usepackage[usenames,dvipsnames]{xcolor}

\usepackage{graphicx}
\usepackage{hyperref}
\newcommand{\ben}{\begin{equation}}
\newcommand{\een}{\end{equation}}
\newcommand{\bea}{\begin{eqnarray}}
\newcommand{\eea}{\end{eqnarray}}

\def\sss{\scriptscriptstyle\rm}

\def\1s{_{1,\sss S}}
\def\2s{_{2,\sss S}}

\def\dulR{{\underline{\underline{\bf R}}}}

\begin{document}
 \title{Energy-Conserving Coupled Trajectory Mixed Quantum Classical Dynamics}
 \author{Evaristo Villaseco Arribas}
 \affiliation{Department of Physics, Rutgers University, Newark 07102, New Jersey USA}
 \author{Neepa T. Maitra}
 \affiliation{Department of Physics, Rutgers University, Newark 07102, New Jersey USA}
 \email{neepa.maitra@rutgers.edu}
 \date{\today}

\begin{abstract}

\begin{minipage}[t]{0.5\linewidth}
The coupled-trajectory mixed quantum classical method (CTMQC), derived from the exact factorization approach,  has successfully predicted photo-chemical dynamics in a number of interesting molecules, capturing population transfer and decoherence from first-principles. However, due to the approximations made, CTMQC does not guarantee energy conservation.  We propose a modified algorithm, CTMQC-E, which redefines the integrated force in the coupled-trajectory term so to restore energy conservation, and demonstrate its accuracy on scattering in Tully's extended coupling region model and photoisomerization in a retinal chromophore model.
\end{minipage}
\vtop{%
  \vskip-1ex
  \hbox{%
    \includegraphics[width=0.22\linewidth]{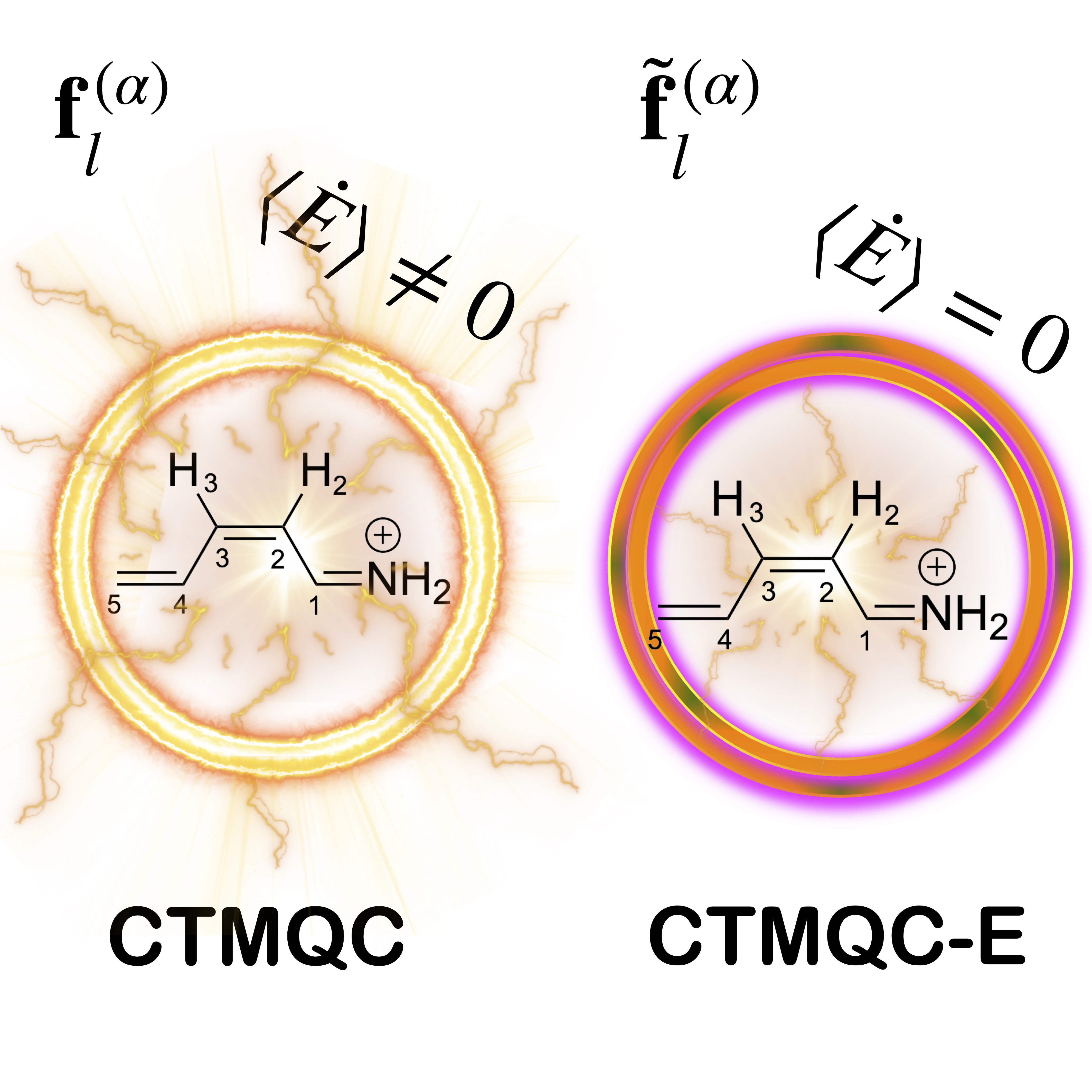}%
  }%
}%
\end{abstract}
\maketitle

The first law of thermodynamics states that the energy of a closed system must be conserved. For molecular systems, this means energy exchange must occur between the nuclei and electrons in such a way that the total energy of the molecule is invariant. To simulate the dynamics of complex molecules approximations are inevitably required, and when energy conservation does not arise naturally in a given approximation, it is usually imposed as an extra condition. In some cases, the energy non-conservation is a consequence of simplifications made in the numerical implementation of the method, e.g. in the independent-trajectory version of the multiconfigurational Ehrenfest method in finite-Gaussian basis set implementations~\cite{MB20,AJI22}. Justified by the large nuclear-electron mass ratio, mixed quantum-classical (MQC) schemes are the basis for many approximations for coupled electron-nuclear dynamics,  where one propagates an ensemble of classical nuclear trajectories, each associated with a set of quantum electronic coefficients of, typically, a Born-Oppenheimer (BO) basis. Ehrenfest dynamics (EH) and surface-hopping (SH)~\cite{T90,T98} are the most widely-used MQC methods, with SH generally preferred due to its ability to capture wavepacket splitting after regions of coupling between BO surfaces are encountered.  Both conserve energy, albeit in different ways. Both involve a nuclear force that is the gradient of an electronic expectation value, with the electronic equations having a Hamiltonian evolution.
But while the consistency in the electronic and nuclear equations in Ehrenfest yields energy conservation on an individual trajectory level, the inconsistency in having coherent electronic evolution while each nuclear trajectory is collapsed on a given surface at any time, means energy conservation  in SH is a more subtle issue. 
%In a mixed quantum classical (MQC) scheme this means that during the dynamics energy exchange can occur between the classical (nuclei) and quantum (electronic) degrees of freedom in such a way that the total energy of the molecule remains invariant.
%Two of the most widely-used mixed quantum classical schemes Surface Hopping (SH)~\cite{T90,T98}.

 SH imposes energy conservation on each individual trajectory by rescaling the velocities after a hop has occurred such that the gain or loss in potential energy is compensated by a loss or gain in kinetic energy. Not only is there no unique way to do this~\cite{CGB17, B21,TSF21}, but the notion itself lacks proper physical justification since it is the energy of the ensemble of trajectories that mimics the underlying quantum wavepacket that should be conserved, not the energy of an individual trajectory, and this too tight constraint leads to inaccuracies. One example is the so-called frustrated hop, where strict energy conservation precludes dynamical pathways that are accessible in the full quantum mechanical treatment. Furthermore, frustrated hops exacerbate the difference between electronic populations predicted from the electronic coefficient evolution and those predicted by the fraction of nuclear trajectories on the corresponding electronic state, which is known as the internal consistency problem of SH. 
More sophisticated SH schemes that do conserve energy over the ensemble rather than for individual trajectories~\cite{M16b,M19} have not been applied beyond model systems so far. But even aside from the incorrect imposition of individual trajectory energy conservation and its {\it ad hoc} manner of imposition, there is a third issue: if the electronic populations in the definition of the energy were obtained from the coherent electronic evolution instead of the fraction of trajectories, then SH methods would in fact not conserve total energy. Although SH schemes that include a decoherence correction~\cite{GP07,WAP16,CB18,SJLP16} ameliorate the internal inconsistency (albeit in an {\it ad hoc} way), the other issues remain.

The exact factorization approach (XF)~\cite{Hunter75,Hunter_IJQC1975_2,Hunter_IJQC1980,H81,Hunter_IJQC1982,GG14,AMG10,AMG12,AG21,VMM22} paved the way for the development of new MQC methods that tackle the issues of standard MQC schemes. The coupled-trajectory mixed quantum classical algorithm (CTMQC)~\cite{MAG15,AMAG16,MATG17} was derived by taking the classical limit of the XF equations and approximating some of the terms in ways justified by exact studies~\cite{AASMMG15}. Electronic decoherence and nuclear wavepacket-splitting emerge naturally, and CTMQC has been  successfully applied to a number of interesting photo-induced dynamics~\cite{MATG17,Tavernelli_EPJB2018,MOLA20,PMLA21,TLA22,GCTMQC}. The XF terms introduce new mechanisms of population transfer driven by nuclear quantum momentum, shown to be crucial in capturing dynamics of multi-state intersections~\cite{VMM22} in an algorithm where the electronic equation was used in a SH scheme~\cite{HLM18,PyUNIxMD,FMK19,FPMK18,FPMC19,FMC19,VIHMCM21,VMM22}.

However, due to the approximations made to the XF equations, the ensemble of coupled trajectories in CTMQC is not guaranteed to conserve the total energy. In this work, we analyze why, and propose and demonstrate a modification, CTMQC-E, that restores energy conservation.

The CTMQC electronic coefficients and nuclear force evolve via
\begin{equation}
\dot{C}_l^{(\alpha)}(t)=\dot{C}_{l,Eh}^{(\alpha)}(t)+\dot{C}_{l,XF}^{(\alpha)}(t)\label{eq:cxf}
\end{equation}
\begin{equation}
\bold{F}_\nu^{(\alpha)}(t)=\bold{F}_{\nu,Eh}^{(\alpha)}(t)+\bold{F}_{\nu,XF}^{(\alpha)}(t)
\label{eq:fxf}
\end{equation}
where the first terms in the electronic and nuclear equations are Ehrenfest-like terms:
\begin{equation}
\dot{C}_{l,Eh}^{(\alpha)}=-i\epsilon_l^{(\alpha)}C_{l}^{(\alpha)}-\sum_{k}\sum_{\nu}^{N_n}{\bold{R}}_{\nu}^{(\alpha)}\cdot\bold{d}_{\nu,lk}^{(\alpha)}C_{k}^{(\alpha)}
\end{equation}
\begin{equation}
\bold{F}_\nu^{(\alpha)}=-\sum_l\rho_{ll}^{(\alpha)}\nabla_{\nu}\epsilon_l^{(\alpha)}+\sum_{l,k}\rho_{lk}^{(\alpha)}\Delta \epsilon_{lk}^{(\alpha)}\bold{d}_{\nu,lk}^{(\alpha)}C_{k}^{(\alpha)}\,.
\end{equation}
Atomic units ($\hbar\,$=$\,m_e\,$=$\,e^2\,$=$\,1$) are used throughout this letter and the shorthand notation $f^{(\alpha)}=f(\dulR^{(\alpha)}(t))$ denotes evaluation at the position of trajectory $\alpha$. Here and henceforth we drop the $t$-dependences of the quantities in writing the equations, to avoid notational clutter.
%The label $(t)$ indicating explicit time-dependence is dropped for simplicity. 
The sum over $\nu$ is a sum over all $N_n$ nuclei.  The  electronic density-matrix elements are $\rho_{lk}^{(\alpha)}=C_l^{(\alpha)*}C_k^{(\alpha)}$,  $\bold{d}_{\nu,lk}^{(\alpha)}$ is the nonadiabatic coupling vector (NAC) along the $\nu$ nuclear coordinate 
between BO states $l$ and $k$ and $\Delta \epsilon_{lk}^{(\alpha)}$ the BO energy difference between states $l$ and $k$; the sums over Latin indices go over the electronic states.   The second terms in Eqs (\ref{eq:cxf}) and (\ref{eq:fxf}) are the corrections coming from XF 
\begin{equation}
\dot{C}_{l,XF}^{(\alpha)}=\sum_\nu^{N_n}\sum_k\frac{\bold{Q}_\nu^{(\alpha)}}{M_\nu}\cdot\Delta\bold{f}_{\nu,lk}^{(\alpha)}\rho^{(\alpha)}_{kk}C_l^{(\alpha)}
\label{eq:el_eom}
\end{equation}
\begin{equation}
\bold{F}_{\nu,XF}^{(\alpha)}(t)=\sum_\mu^{N_n}\sum_{l,k}\frac{2\bold{Q}_\mu^{(\alpha)}}{M_\mu}\cdot\bold{f}_{\mu,l}^{(\alpha)}\rho^{(\alpha)}_{ll}\rho^{(\alpha)}_{kk}\Delta\bold{f}_{\nu,lk}^{(\alpha)}
\label{eq:nuc_force}
\end{equation}
with $\Delta\bold{f}_{\nu,lk}^{(\alpha)}=\bold{f}_{\nu,l}^{(\alpha)}-\bold{f}_{\nu,k}^{(\alpha)}$ where $\bold{f}_{\nu,lk}^{(\alpha)}$ is the time-integrated adiabatic force on nucleus $\nu$ accumulated on the $l$-th surface along the trajectory $\alpha$ 
\begin{equation}
\bold{f}_{\nu,l}^{(\alpha)}=-\int_0^t\nabla_\nu E_l^{(\alpha)}d\tau\,,
\label{eq:af}
\end{equation}
and $\bold{Q}_\nu^{(\alpha)}$  the nuclear quantum momentum evaluated at the position of the trajectory $\dulR^{(\alpha)}(t)$
\begin{equation}
\bold{Q}_\nu^{(\alpha)}=-\frac{\nabla_\nu\vert\chi^{(\alpha)}\vert^2}{2\vert\chi^{(\alpha)}\vert^2}
\label{eq:qmom}
\end{equation}

Variations of the CTMQC algorithm have been explored, for example, using the electronic equation~Eq.(\ref{eq:cxf})  within a SH or Ehrenfest scheme~\cite{HLM18,PA21,HM22}.
%and EH~\cite{HM22} schemes, resulting in a family of XF-based MQC algorithms which have been successfully applied to a variety of systems~\cite{MATG17, Tavernelli_EPJB2018,MOLA20,FMK19,FPMK18,FPMC19,FMC19,PA21,VIHMCM21,VMM22}. 
A central term in the XF-based MQC methods is %the coupling of trajectories through 
the nuclear quantum momentum Eq.~(\ref{eq:qmom}). Although the SH scheme of Ref.~\cite{HLM18} computes this with auxiliary trajectories in order to yield an independent-trajectory method, and the corresponding independent-trajectory version of CTMQC was recently implemented~\cite{HM22}, a key feature of the original CTMQC algorithm is the coupling of trajectories through this term. The nuclear quantum momentum can be computed in two different ways. The original definition (Q$_0$) implies using expression Eq.(\ref{eq:qmom}) reconstructing the nuclear density  as a sum of gaussians centered at the position of the classical trajectories. However, this definition was found to sometimes yield an unphysical net population transfer in regions of vanishing NAC. Instead, the modified definition (Q$_m$) was constructed by requiring, zero net population transfer in regions of zero NAC~\cite{AMAG16,MATG17}. The condition is imposed pairwise, and separately for       each degree of freedom, resulting in
\begin{equation}
\sum_\alpha^{N_{tr}}\frac{Q_{\mu,lk}^{(\alpha)}}{M_\mu}\rho_{ll}^{(\alpha)}\rho_{kk}^{(\alpha)}\Delta f_{\mu,lk}^{(\alpha)}=0\label{eq:qmommodif}
\end{equation}
%Using  Q$_m$ fixes the spurious population transfer that might occur when Q$_0$ is used. 
and this is what has been used in the CTMQC calculations on photo-induced molecular dynamics~\cite{MATG17,Tavernelli_EPJB2018,MOLA20,PMLA21,TLA22}.
A deeper analysis on the different ways of computing $\bold{Q}_\nu^{(\alpha)}$ and its impact on the dynamics of model systems was studied on Ref~\cite{VAM22}.

The condition of zero net population transfer in regions of zero NAC may be viewed as an exact condition, much like energy conservation. The total energy of the molecule is the expectation value of $H$, i.e, the sum of the nuclear kinetic energy and the BO Hamiltonian, $H = T_n + H_{\rm BO}$. For MQC methods, the expectation value involves a nuclear-trajectory average, and leads to the following definition of the trajectory-averaged energy in CTMQC 
\begin{eqnarray}
%\langle E_{CT}\rangle&=& \sum_\alpha^{N_{tr}}  \\
\langle E_{CT}\rangle&=&\frac{1}{N_{tr}}\sum_\alpha^{N_{tr}}\left(\frac{1}{2}\sum_\nu^{N_n} M_\nu\dot{\bold{R}}_\nu^{(\alpha)2}+\sum_l\rho_{ll}^{(\alpha)}\epsilon_l^{(\alpha)} \right)
\label{eq:Ect}
\end{eqnarray}
We note that, compared to a fully quantum scheme, this definition misses a contribution to the nuclear kinetic energy from the second-derivative of the nuclear density, which is higher order in a semiclassical expansion and neglected in deriving the CTMQC equations of motion for the trajectories~\cite{AMAG16}, however the definition Eq.~(\ref{eq:Ect}) is consistent with the MQC framework. 

Taking the time-derivative of the energy yields
\begin{eqnarray}
\nonumber
\langle \dot{E}_{CT}\rangle&=&\frac{1}{N_{tr}}\sum_\alpha^{N_{tr}}\left(\bold{F}_{\nu,XF}^{(\alpha)}\cdot\dot{\bold{R}}_\nu^{(\alpha)}+\sum_l^{N_{st}}\dot{\rho}_{ll,XF}^{(\alpha)}\epsilon_l^{(\alpha)}\right)\\
%\begin{eqnarray}
&=&\frac{1}{N_{tr}}\sum_\alpha^{N_{tr}}\sum_\mu\frac{\bold{Q}_\mu^{(\alpha)}}{M_\mu}\cdot\sum_{l,k}\rho_{ll}^{(\alpha)}\rho_{kk}^{(\alpha)}\Delta\bold{f}_{\mu,lk}^{(\alpha)}\nonumber \\
&&\times\Bigg [\sum_\nu\Delta\bold{f}_{\nu,lk}^{(\alpha)}\cdot\dot{\bold{R}}_\nu^{(\alpha)}+\Delta\epsilon_{lk}^{(\alpha)}\Bigg] \label{eq:ec}
\end{eqnarray}
Eq.~(\ref{eq:ec}) shows that, in general, energy-conservation is not guaranteed in CTMQC, which may lead to unphysical dynamics of electronic and/or nuclear observables.
%The XF terms which are essential to capture quantum (de)coherence  leads on the other to breaking of energy conservation ($\langle \dot{E}_{CT}\rangle\neq0$). 
%This could lead to the wrong population dynamics in some scenarios.

Inspired by how the quantum momentum was modified to impose the exact condition of zero net population transfer in regions of zero NAC~\cite{AMAG16,MATG17} (Eq.~\ref{eq:qmommodif}), here we propose a modified definition of the accumulated force $\bold{f}_{\mu,l}^{(\alpha)}$ that satisfies energy conservation. Applying the energy condition through redefining $\bold{f}_{\mu,l}^{(\alpha)}$ was inspired by Ref.~\cite{HM22} which fixed the energy conservation in the independent-trajectory version of CTMQC that uses auxiliary trajectories.
%s by proposing an alternative approximation for $\bold{f}_{\mu,l}^{(\alpha)}$. 
%that guaranteed energy conservation at the individual trajectory level. 
As noted earlier, this concept of energy-conservation is likely too strict, and it should instead be the energy of the {\it ensemble} that is conserved.  First,  in the framework of the modified definition of the quantum momentum Eq.~(\ref{eq:qmommodif}), we see from  Eq.~(\ref{eq:ec}) that energy will be conserved in situations where the quantity in the square brackets in the last line,
\begin{equation}
\sum_\nu\Delta\bold{f}_{\nu,lk}^{(\alpha)}\cdot\dot{\bold{R}}_\nu^{(\alpha)}+\Delta\epsilon_{lk}^{(\alpha)}
\label{eq:temp}
\end{equation}
is trajectory-independent.  We enforce this by setting Eq.~(\ref{eq:temp}) equal to its trajectory average. For one degree of freedom this means
\begin{equation}
\Delta\tilde{f}_{lk}^{(\alpha)}\dot{R}^{(\alpha)}+\Delta\epsilon_{lk}^{(\alpha)}= \frac{1}{N_{tr}}\sum_\alpha \left(\Delta f_{lk}^{(\alpha)}\dot{R}^{(\alpha)}+\Delta\epsilon_{lk}^{(\alpha)}\right)
\end{equation}
That is, for one degree of freedom, the accumulated force is redefined as
\begin{equation}
\tilde{f}_{l}^{(\beta)}=\frac{1}{\dot{R}^{(\beta)}}\left[-\epsilon_l^{(\beta)}+\frac{1}{N_{tr}}\sum_\alpha f_l^{(\alpha)} \dot{R}^{(\alpha)}+\epsilon_l^{(\alpha)}\right]
\label{eq:afs}
\end{equation}
For multiple degrees of freedom, we require 
\begin{equation}
\sum_\nu\tilde{\bold{f}}_{\nu,l}^{(\beta)}\cdot\dot{\bold{R}}_\nu^{(\beta)}+\epsilon_l^{(\beta)} = \frac{1}{N_{tr}}\sum_{\alpha\nu}\bold{f}_{\nu,l}^{(\alpha)}\cdot\dot{\bold{R}}_\nu^{(\alpha)}+\epsilon_l^{(\alpha)}\,,
\end{equation}
which can be satisfied by choosing
\begin{equation}
\tilde{\bold{f}}_{\nu,l}^{(\beta)}=\frac{ \left(-\epsilon_l^{(\beta)}+\frac{1}{N_{tr}}\sum_{\alpha\nu} \bold{f}_{\nu,l}^{(\alpha)}\cdot\dot{\bold{R}}_\nu^{(\alpha)}+\epsilon_l^{(\alpha)}\right)}{\sum_\nu\bold{n}_\nu^{(\beta)}\cdot\dot{\bold{R}}_\nu^{(\beta)} }\bold{n}_\nu^{(\beta)}
\end{equation}
where $\bold{n}_\nu^{(\beta)}$ is an arbitrary vector defining the direction of $\tilde{\bold{f}}_{\nu,l}^{(\beta)}$. Similar to the fix in the independent-trajectory version of CTMQC~\cite{HM22}, we choose   $\bold{n}_{\nu}^{(\beta)}= M_\nu\dot{\bold{R}}^{(\beta)}_\nu$ as the momentum of the trajectory, so that
\begin{equation}
\tilde{\bold{f}}_{\nu,l}^{(\beta)}=\frac{ \left(-\epsilon_l^{(\beta)}+\frac{1}{N_{tr}}\sum_{\alpha\nu} \bold{f}_{\nu,l}^{(\alpha)}\cdot\dot{\bold{R}}_\nu^{(\alpha)}+\epsilon_l^{(\alpha)}\right)}{2E_{kin}^{(\beta)}/M_\nu }\dot{\bold{R}}_\nu^{(\beta)}
\label{eq:newf}
\end{equation}
%As it was noted in~\cite{HM22} this makes the XF-force behave like a friction. Iam not so sure, a friction force would be prop to the veloc but this is integrated force so is actually more a momentum quantity...maybe similar though...

Eq.~(\ref{eq:newf}) is the main result of this paper, and when this is used in the CTMQC algorithm, we call the resulting method CTMQC-E. CTMQC-E is relatively straightforward to implement on top of an existing CTMQC code, such as g-CTMQC~\cite{GCTMQC}, which we have used in the following work to test CTMQC-E on two different systems. 

\begin{figure}[h!]
 \begin{center}
\includegraphics[width=0.5\textwidth]{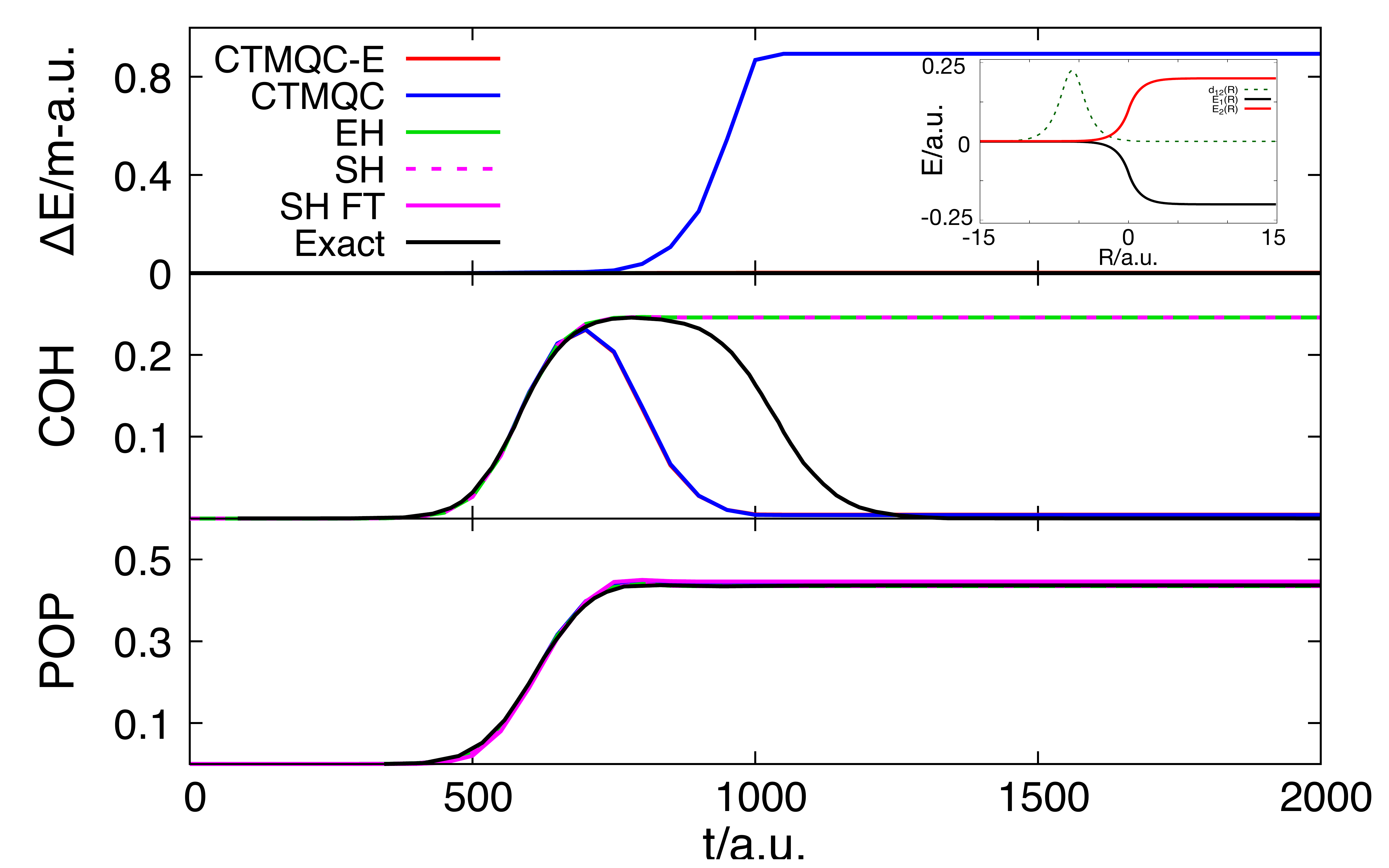}
\caption{Total energy deviation ($E(t) - E(0)$) in miliatomic units (upper panel), coherences (middle panel) and excited state populations (lower panel) as a function of time for ECR model with k$_0$=32 a.u. The inset in the upper panel shows the two BO surfaces in solid lines and the NAC in dashed, as a function of $R$}.
\label{fig:k32}
\end{center}
\end{figure}

\begin{figure}[h!]
 \begin{center}
\includegraphics[width=0.5\textwidth]{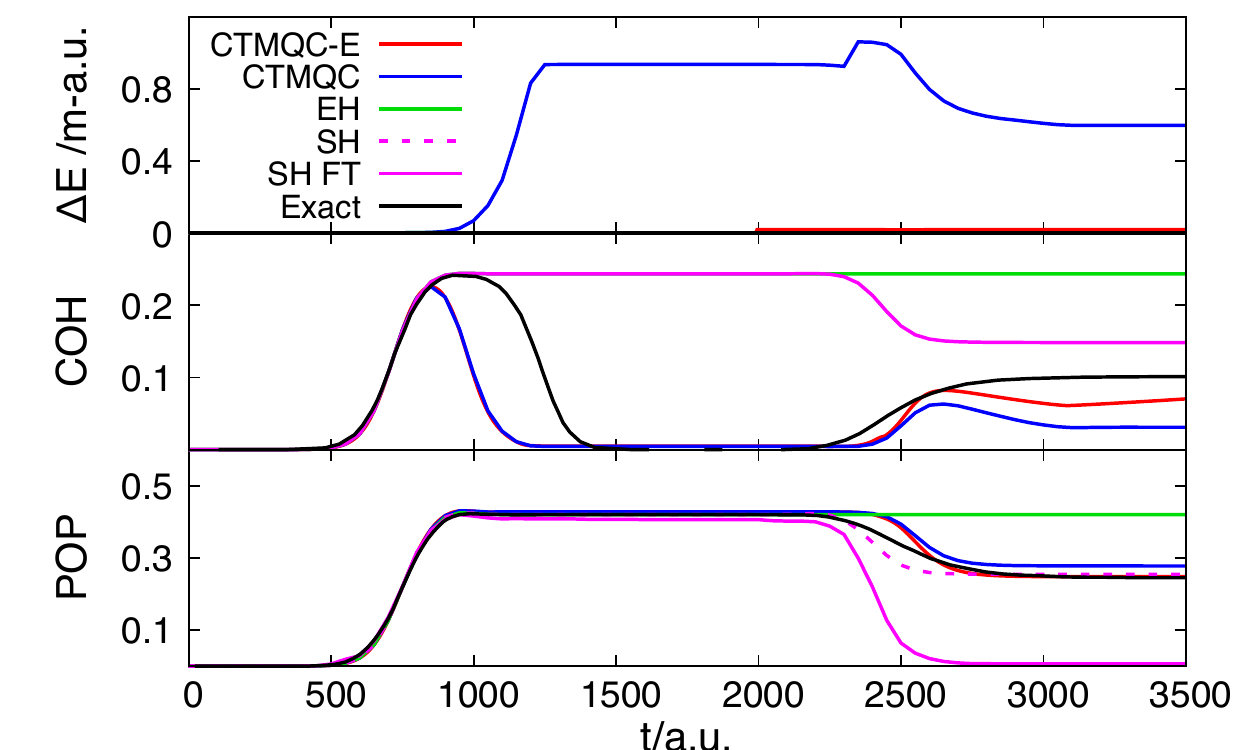}
\caption{Total energy deviation in miliatomic units (upper panel), coherences (middle panel) and excited state populations (lower panel) as a function of time for ECR model with k$_0$=26 a.u.}
\label{fig:k26}
\end{center}
\end{figure}

 %We assessed the performance of CTMQC-E for one and several degrees of freedom. The code used for MQC dynamics is a modified, developer version of the g-CTMQC code\cite{GCTMQC}. 
Our first example is the one-dimensional Tully's extended coupling region (ECR) model~\cite{T90}, illustrated in the inset of Fig.~\ref{fig:k32}. We study two situations where a  Gaussian nuclear wavepacket centered at $-15$ a.u. is incoming on the lower surface from the left with a high  ($k_0$ = 32 a.u.) and a low ($k_0$ = 26 a.u.) initial momentum; the variance of the gaussian wavepacket at time zero 20 times the inverse of the initial momentum. These two scenarios were studied with independent-trajectory XF-based methods in Ref.~\cite{HM22}. For the higher initial momentum case after passing through the NAC region where some population is transferred to the upper surface, the wavepacket component on the lower surface moves faster separating in nuclear space from the part that has transferred to the upper state and the branches decohere. For the lower initial momentum, much of the wavepacket transmitted to the upper surface reflects back and recrosses the NAC region leading to a second splitting event. 
%Ref~\cite{HM22} simulated these two scenarios with two independent-trajectory XF-based MQC schemes: a modified SH scheme (SHXF) and the independent trajectory version of CTMQC (MQCXF). They studied the impact of employing two different auxiliary propagation schemes, of making the width of the auxiliary gaussian time-dependent and of enforcing energy conservation for each trajectory.
In our MQC simulations $1000$ Wigner-distributed trajectories with a fixed initial momentum $k_0$ were run starting on the lower surface and the time-step used in the calculations was 0.1 a.u. 
Fig.~\ref{fig:k32} and Fig.~\ref{fig:k26} show the excited state population $\sum_\alpha \rho_{22}^{(\alpha)}$, the fraction of trajectories on the active state for SH $N_2(t)/N_{tr}$, the coherence $\sum_\alpha \vert\rho_{12}^{(\alpha)}\vert^2$, and energy from Eq.~(\ref{eq:Ect}) for $k = 32$a.u. and $k = 26$ a.u. respectively.

Fig.~\ref{fig:k32} shows that for $k_0 = 32$a.u. momentum all MQC methods closely reproduce the exact populations. CTMQC and CTMQC-E give identical coherences that decay faster than the exact, whereas SH and EH do not decohere. The top panel shows that CTMQC-E conserves the total classical energy, curing the increase in CTMQC between 500 and 1000 a.u. where the quantum momentum is active. SH and EH conserve energy so are not shown in the top panel. After 1000 a.u. the trajectories fully decohere and energy thereafter remains constant in both methods.
These results could be compared with the independent trajectory XF-based schemes shown in Fig. 5 of Ref.~\cite{HM22}. There, the populations and coherences are also closely reproduced, but there are some variations, depending on the way the width parameter of the auxiliary trajectory is computed, and the equations are integrated. 

%With MQCXF in all cases there is some spurious back-transfer to the ground state inherent to the quantum momentum computed with auxiliary trajectories~\cite{VAM22}. Interestingly 
%the problem is exacerbated by using a time-dependent width and energy-conserving phase.

Fig.~\ref{fig:k26} shows the $k_0 = 26$a.u. case.  Consider first the standard MQC schemes. EH does not capture any reflection or decoherence after the first passage through the NAC region. On the other hand, the fraction of trajectories population measure of SH predicts almost complete reflection of all trajectories on the upper surface distinct from the behavior of its electronic populations, displaying poor internal consistency. As discussed earlier, if we were to take the electronic populations as the population observable, SH would not conserve energy.
Now let's turn to the XF-based MQC methods. CTMQC-E gives an improvement over CTMQC in the populations, coherence, and total energy. We note that to avoid numerical instabilities occurring when the velocity becomes too small in the denominator of Eq.~(\ref{eq:afs}) (an issue for reflecting trajectories), we apply a velocity threshold such that the new definition of accumulated force from Eq.~(\ref{eq:afs}) is used only above a small threshold ($10^{-5}$), and otherwise the original definition Eq.~(\ref{eq:af}) is used. This procedure can lead to small discontinuities in the energy, as seen in the red curve in Fig~\ref{fig:k26} at around 2000 a.u. when the energy jumps up a little above the initial value. Further work will involve stabilizing the algorithm, but we see already the  improvement in energy-conservation in CTMQC-E over CTMQC. Further, the populations are slightly improved, and also the coherence. 
These results may again be compared with the independent trajectory XF-based schemes in Fig. 4 of Ref.~\cite{HM22}, where the results were quite sensitive to details in the algorithm, and a time-dependent auxiliary-trajectory width was needed to capture the exact dynamics in the energy-conserving scheme. 

Our second example is a three-dimensional model of the the photo-induced isomerization of the 2-cis-penta-2,4-dieniminium cation (PSB3), which in turn is a model of the retinal cromophore of Rhodopsin (rPSB11), responsible for dim-light vision in vertebrates. This was extensively studied in Ref.~\cite{MOLA20}, by means of quantum and mixed quantum-classical dynamics.
%\begin{figure}[h!]
% \begin{center}
%\includegraphics[width=0.2\textwidth]{psb3}
%\caption{2-cis-penta-2,4-dieniminium cation (PSB3)}
%\label{fig:psb3}
%\end{center}
%\end{figure} 
The reduced-dimensionality model is based on 3 degrees of freedom ($r_B$, $\theta_T$, $\phi_H$) developed in Ref~\cite{MFYVO19}. These are the bond-length-alternation stretching (BLA) defined as the average length difference between single and double bonds
\begin{equation}
r_{B}=\frac{d_{C_1C_2}+d_{C_3C_4}}{2}-\frac{d_{NC_1}+d_{C_2C_3}+d_{C_4C_5}}{3}   \;, 
\end{equation}
the torsional deformation around the double reactive bond C$_2$=C$_3$ (TORS)
\begin{equation}
\theta_{T}=dihedral(\mathrm{C}_1\mathrm{C}_2\mathrm{C}_3\mathrm{C}_4)  \;,
\end{equation}
and the hydrogen-out-of-plane wagging (HOOP) defined as the difference between the TORS and $\mathrm{H}_2\mathrm{C}_2\mathrm{C}_3\mathrm{H}_3$ dihedral angles
\begin{equation}
\phi_{H}=\theta_T-dihedral(\mathrm{H}_2\mathrm{C}_2\mathrm{C}_3\mathrm{H}_3)    
\end{equation}

These three modes which drive the ultrafast cis-trans isomerization of rPSB11~\cite{SRFF11,KWGS12,EMNWOG18,CFJK18} also drive the isomerization of cis-PSB3 after photo-excitation from the ground (S$_0$) to the first singlet excited state (S$_1$)~\cite{GHSLGAO12,GMLKGACO13}. We propagated $600$ Wigner-distributed trajectories starting in the S$_1$ state, with centers at 0.154449 a.u. for the BLA coordinate and at zero angle for TORS and HOOP, all with zero center momentum. The variances were 0.154449 a.u. for BLA, 0.183302 for TORS, and 0.406143 for HOOP. 
%initial values of the position, momenta, and the variances of the gaussian wavepackets in each dimension at time zero are gathered in Table~\ref{table:1}. {\color{magenta} i don't think we need the table, it takes too much space; we can describe easily, esp since many are 0 or 1...}
%\begin{table}[h!]
%\begin{tabular}{c|c|c|c|c|}
%\cline{2-5}
%                           & $R_0$ & $K_0$ & %$\sigma_{0,R}$ & $\sigma_{0,k}$ \\ \hline
%\multicolumn{1}{|l|}{BLA}  & 0.154449 & 0.0 & 0.154449 & -1.0    \\ \hline
%\multicolumn{1}{|l|}{TORS} & 0.0 & 0.0 & 0.183302 & -1.0 \\ \hline
%\multicolumn{1}{|l|}{HOOP} & 0.0 & 0.0 & 0.406143 & -1.0 \\ \hline
%\end{tabular}
%\caption{Centers and variances of the gaussian wavepackets in each dimension at time zero.}
%\label{table:1}
%\end{table}
We used a time-step of 0.01a.u;  a larger time-step gave qualitatively similar results but, for this system, led to larger energy-conservation violation in CTMQC and also larger jumps in CTMQC-E. Note that Ref.~\cite{MOLA20} dealt with the energy non-conservation simply by reverting to Ehrenfest dynamics after 120 fs.

\begin{figure}[h!]
 \begin{center}
\includegraphics[width=0.45\textwidth]{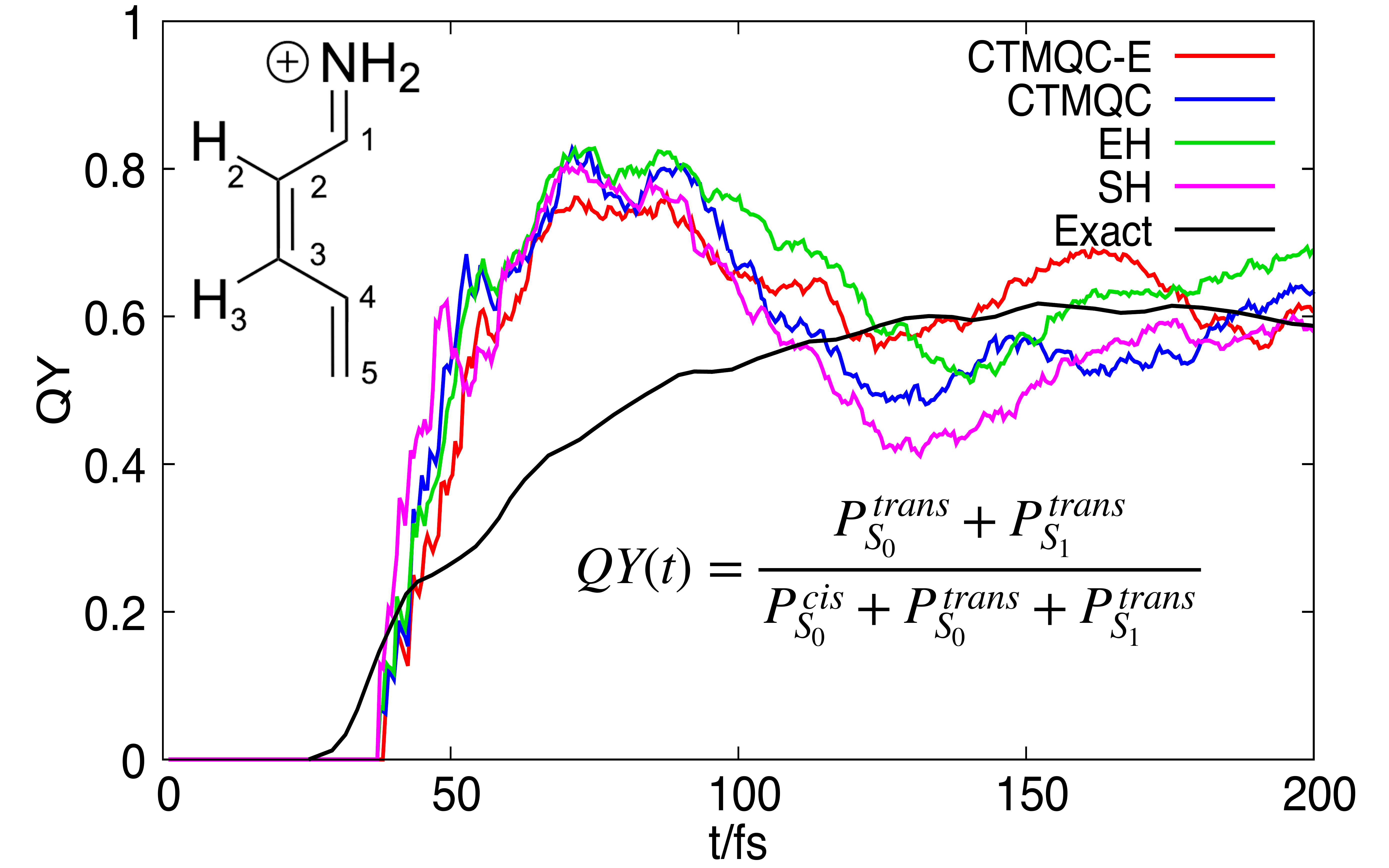}
\caption{Quantum yield of the PBS3 photo-isomerization as a function of time.}
\label{fig:qy}
\end{center}
\end{figure}

Fig.~\ref{fig:qy} shows the quantum yield of the photo-isomerization process, defined as the ratio between the trans and all reaction products, as indicated in the figure. All methods overestimate the quantum yield at short times; Ref.~\cite{MOLA20} attributed this to the lack of nuclear quantum effects. 
%due to {\color{red} trajectories funneling too fast through the S$_0$/S$_1$ connical intersections (CIs) and acquiring a large amount of kinetic energy along the low-frequency TORS mode}.  
After about 120 fs CTMQC-E results oscillate around the exact quantum yield, giving a small improvement over CTMQC which shows an underestimate, and over EH and SH.

\begin{figure}[h!]
 \begin{center}
\includegraphics[width=0.45\textwidth]{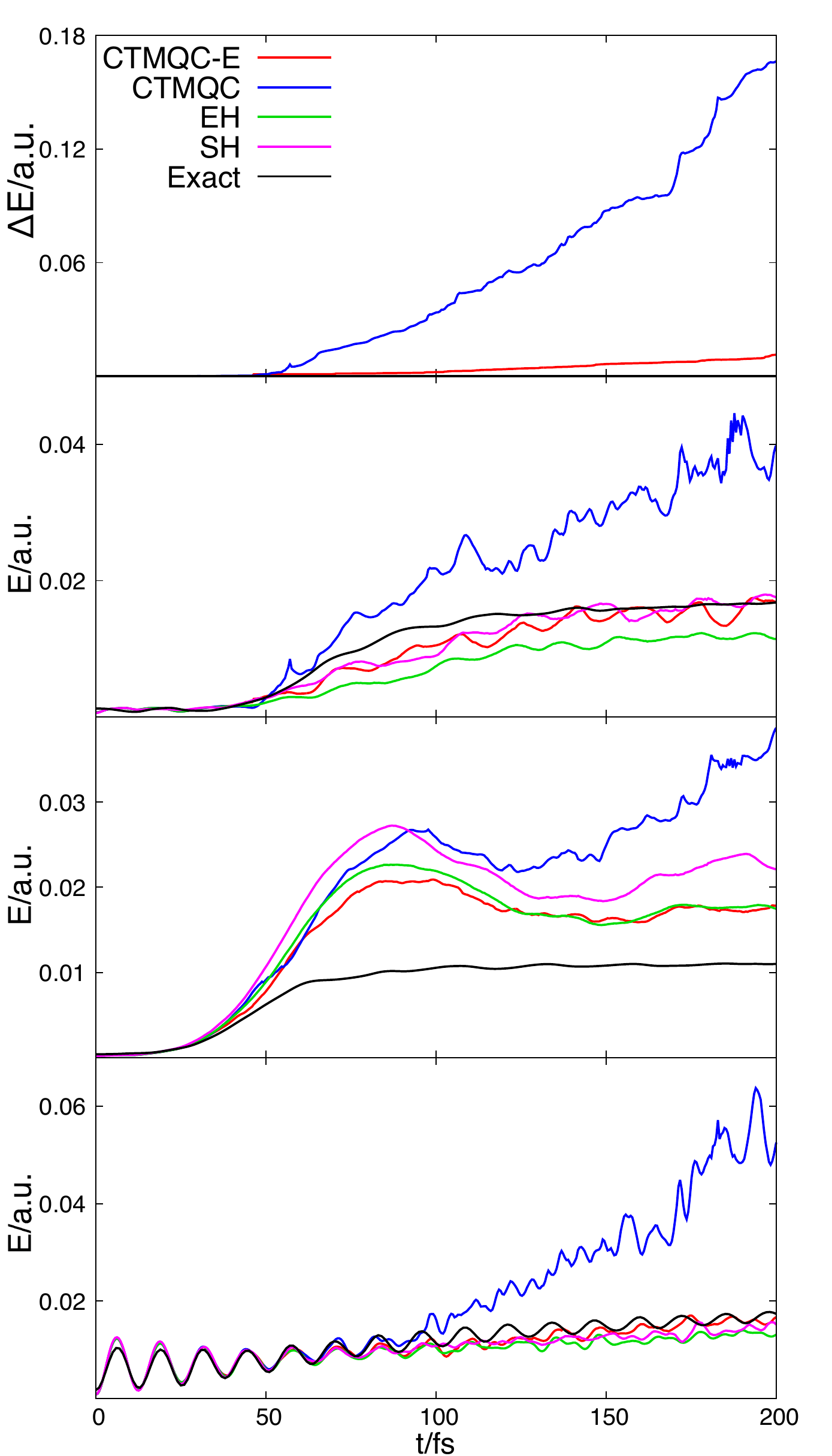}
\caption{Trajectory-averaged total energy deviation for PSB3 as a function of time (top panel);  kinetic energy along the TORS (second from top panel), HOOP (third panel), and BLA (lowest panel) degrees of freedom.}
\label{fig:ekins}
\end{center}
\end{figure}

The top panel in Fig.~\ref{fig:ekins} shows the trajectory-averaged total energy as well as the nuclear kinetic energies  along the BLA , TORS and HOOP modes, as compared with EH and SH dynamics and exact calculations. 
Although not entirely, CTMQC-E strikingly reduces the total energy violation of CTMQC. The residual deviation from constant energy in CTMQC-E is caused by two effects. First, for trajectories for which the denominator in Eq.~(\ref{eq:newf}) becomes smaller than a fixed threshold, the accumulated force reverts to its original definition, which does not conserve energy. Second, even when the new definition is used, the quantum momentum computation may revert to the original quantum momentum Q$_0$ definition when a denominator involved in imposing Eq.~(\ref{eq:qmommodif}) becomes too small~\cite{MATG17,VAM22}; our redefined force is guaranteed to conserve energy only when used in conjunction with the modified definition Q$_m$. 
Considering now the kinetic energy along BLA (lowest panel), in the exact system it oscillates around an average that increases in a monotonic way due to the wavepacket moving through the conical intersections towards S$_0$, while the oscillations decrease in amplitude as a result of a loss in vibrational coherence as the wavepacket spreads along the TORS degree of freedom.  CTMQC captures the initial behavior well but starts increasing at around 80 fs and deviating significantly from the exact calculations; EH and SH, in contrast, over-damp the BLA oscillations and average. 
%Fig.~\ref{fig:its} show the trajectory-averaged value of the BLA coordinate as a function of time obtained from CTMQC, CTMQC-E and EH dynamics compared with the expectation value obtained from the exact results.
%As we can see after 100 fs oscillations in the BLA coordinate in EH get dampened too much. On the other hand in CTMQC due to energy violation the BLA coordinate gains kinetic energy and starts oscillating with large amplitudes. 
 On the other hand, CTMQC-E reproduces the exact kinetic energy along BLA reasonably.
For the torsional kinetic energy  all MQC schemes yield an overestimate of the kinetic energy from around 40 fs. In CTMQC-E, EH and SH, this gain in TORS kinetic energy is compensated by a loss in HOOP kinetic energy and a loss in potential energy (not shown here) compared with the exact results. On the other hand in CTMQC,  the kinetic energies of all three degrees of freedom increase simultaneously after 120 fs as a consequence of energy violation. This example appears to be among the most challenging so far for CTMQC for energy conservation, and further work is underway to study this in more detail. 

%\begin{figure}[h!]
% \begin{center}
%\includegraphics[width=0.5\textwidth]{BLA-599.pdf}
%\includegraphics[width=0.5\textwidth]{BLA.pdf}
%\caption{Average value of the BLA coordinate as %function of time.}
%\label{fig:its}
%\end{center}
%\end{figure} 

%\begin{figure}[h!]
 %\begin{center}
%\includegraphics[width=0.5\textwidth]{ekin-epot.pdf}
%\caption{Trajectory averaged potential and kinetic energies for PSB3.}
%\label{fig:ener}
%\end{center}
%\end{figure} 
%\vspace{-0.3cm}

In summary, we have presented a modification of the CTMQC algorithm, CTMQC-E, that imposes energy-conservation over the ensemble of  nuclear trajectories through a redefinition of the accumulated force. Much like the modified definition of the quantum momentum that imposes the exact condition of zero population transfer in regions of zero NAC, the new definition involves all trajectories in its computation. It has similarities in form to that used in the independent-trajectory version of CTMQC, but results in a more  physical condition, being imposed on the ensemble of trajectories rather than on the individual trajectories. 
Future work includes refinement of the algorithmic implementation to better deal with instabilities arising from  small velocities in the denominator of the new definition of the accumulated force and to reduce the jumps occurring in the energy without going to very small time-steps which may be untenable for larger systems. Generally, for most of systems where CTMQC has been applied energy violation is often too small to have much of an effect on its dynamics or physical observables. However as shown here, this is not guaranteed, and future work will also attempt to characterize situations where CTMQC is expected to yield significant energy non-conservation, for which the use of CTMQC-E gives significant improvement.

%where CTMQC's breaking of energy conservation might be important and enforcing this physical constraint needed to capture the dynamics accurately.

\acknowledgements{
We thank Dr. Federica Agostini , Dr. David Lauvergnat and Dr. Lea M. Ibele for helpful discussions.
Financial support from the National Science Foundation Award CHE-2154829 (EVA) and from the Department
of Energy, Office of Basic Energy Sciences, Division of Chemical
Sciences, Geosciences and Biosciences under Award No. DESC0020044 (N.T.M.) and the Computational Chemistry
Center: Chemistry in Solution and at Interfaces funded by the
U.S. Department of Energy, Office of Science Basic Energy
Sciences, under Award DE-SC0019394 as
part of the Computational Chemical Sciences Program is gratefully acknowledged.}
\bibliography{ref_na.bib}

\end{document}